\title{How to produce discreet Gaussian sequences: Algorithm and code}
\author{
Sparisoma Viridi*\\
Nuclear Physics and Biophysics Research Division\\
Institut Teknologi Bandung, Bandung 40132, Indonesia\\
\and
Veinardi Suendo\\
Physical Chemistry Research Division\\
Institut Teknologi Bandung, Bandung 40132, Indonesia\\
\and
*dudung@fi.itb.ac.id
}
\date{\today}

\documentclass[12pt]{article}
\usepackage{graphicx}
\usepackage{amssymb}
\usepackage{hyperref}

\begin{document}
\maketitle

\begin{abstract}
Algorithm and code to produce sequences whose members obey Gaussian distribution function is reported. Discreet and limited number of groups are defined in the distribution function, where each group is represented only with one value instead of a range of value. The produced sequences are also checked back whether they still fit the discreet distribution function. Increasing of number of particles $N$ increases the value of correlation coefficient $R^2$, but increasing number of groups $M$ reduces it. Value $R^2 = 1$ can be found for $N = 1000000$ at least with $M = 5000$ and for $M = 10$ at least with $N = 1000$.
\end{abstract}

\section{Introduction}

Gaussian distribution function plays important role in many fields of science, such as in mathematical modeling \cite{Gershenfeld_2002}, in physical sciences \cite{Boas_1983}, in quantum chemistry \cite{Atkins_1970}, with integral in nuclear physics \cite{Duderstadt_1976}, and in semiconductor devices \cite{Sze_1985}. Then a need comes up how a sequence, that its members obey Gaussion distribution function, could be produced, since it is needed, for example in molecular dynamics simulatons \cite{Sustini_2011}. A procedure to produce the sequences is presented in algorithm and C++ code.

\section{Gaussian distribution function}

Gaussian or normal distribution function can be represented in the form of

\begin{equation}
\label{eq01}
f(z) = \frac{1}{\sigma\sqrt{2\pi}} \exp\left[-\frac{(z - \mu)^2}{2\sigma^2}\right]
\end{equation}

\noindent
where $\mu$ is the average of $z$ and $\sigma$ is the width of normal distribution curve. The factor in front of right side of Equation (\ref{eq01}) is due to normalization of $f(z)$ integral

\begin{equation}
\label{eq02}
\int_{-\infty}^{\infty} f(z) dz = 1.
\end{equation}

\noindent
Variable $z$ is a certain parameter that obeys Gaussian distribution function, it can be particle velocity, particle diameter, or particle mass.

\subsection{Proof of normalization}

Equation (\ref{eq02}) can be proved using

\begin{equation}
\label{eq03}
\int_{-\infty}^{\infty} \exp{(-x^2)} dx = \sqrt{\pi},
\end{equation}

\noindent
so that

\begin{eqnarray}
\nonumber
\int_{-\infty}^{\infty} f(z) dz = \int_{-\infty}^{\infty} \frac{1}{\sigma\sqrt{2\pi}} \exp\left[-\frac{(z - \mu)^2}{2\sigma^2}\right] dz \\
\nonumber
= \frac{1}{\sigma\sqrt{2\pi}} \int_{-\infty}^{\infty} \sigma\sqrt{2} \exp\left[-\frac{(z - \mu)^2}{2\sigma^2}\right] d\left(\frac{z - \mu}{\sigma\sqrt{2}}\right) \\
\nonumber
= \left(\frac{1}{\sigma\sqrt{2\pi}}\right)\left(\sigma\sqrt{2}\right) \int_{-\infty}^{\infty} \exp\left[-\frac{(z - \mu)^2}{2\sigma^2}\right] d\left(\frac{z - \mu}{\sigma\sqrt{2}}\right) \\
\nonumber
= \left(\frac{1}{\sigma\sqrt{2\pi}}\right)\left(\sigma\sqrt{2}\right) \left(\sqrt{\pi}\right) = 1.
\end{eqnarray}

\subsection{Meaning of $\mu$ and $\sigma$}

Peak of $f(z)$ is located at $z = \mu$ with value

\begin{equation}
\label{eq04}
f_{\max}(z) = f(\mu) = \frac{1}{\sigma\sqrt{2\pi}}
\end{equation}

\noindent
and at $z = \mu \pm \frac12 \sigma$ it gives

\begin{equation}
\label{eq05}
f\left(\mu \pm \frac12 \sigma\right) = \frac{1}{\sigma\sqrt{2\pi}} \exp\left[-\frac{1}{8}\right] \approx \frac{0.8825}{\sigma\sqrt{2\pi}}.
\end{equation}

\begin{figure}[h]
\centering
\includegraphics[width=9cm]{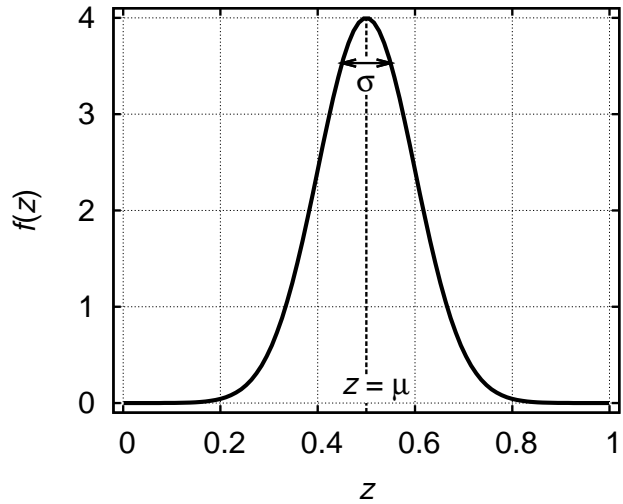}
\caption{\label{fg01} An Gaussian distribution function with $\mu = 0.5$ and $\sigma = 0.25 / \sqrt{2\pi}$.}
\end{figure}

\noindent
An example of $f(z)$ is given in Figure \ref{fg01}.

\subsection{Number of particles}

Suppose that there are $N$ particles in a system, that number of particles $N(z)$ who has property of $z$ is defined by

\begin{equation}
\label{eq06}
N(z) = \frac{N}{\sigma\sqrt{2\pi}} \exp\left[-\frac{(z - \mu)^2}{2\sigma^2}\right],
\end{equation}

\noindent
where according to Equation (\ref{eq02}) it must hold that

\begin{equation}
\label{eq07}
\int_0^{\infty} N(z) dz = N.
\end{equation}

\noindent
In this case, it is considered that property $z$ has only positive value.

\section{Discretization of distribution function}

It is imposible even with nowadays most advanced computer facilities to produce continue number of particles in order of one mole, which equals to about $10^{23}$ particles. In this report only small number of particles is considered. The distribution function is also simplified by dividing it into limited and discreet groups of particles. Within each group there is only one value (certain property of particle) which represents the group instead of a range of value from minimum to maximum value of the group.

\subsection{Discreet groups}

Suppose that there is $M$ groups of particles with equal width $\Delta z$, which group the total number of particles $N$. First step is how to find $z_{\min}$ and $z_{\max}$ where at these values $N(z)$ can be considered zero. Since we deal with particles than it is more simple to use the {\tt int()} function which returns the integer value of $N(z)$. It means that $f(z)$ is considered zero when $N(z) = 1 - \epsilon$, then

\begin{eqnarray}
\label{eq08}
N(z) = 1 - \epsilon, ~z < \mu \Rightarrow z = z_{\rm \min}, \\
\label{eq09}
N(z) = 1 - \epsilon, ~z > \mu \Rightarrow z = z_{\rm \max},
\end{eqnarray}

\noindent
with $\epsilon$ a small defined value. Then width $\Delta z$ can be found through

\begin{equation}
\label{eq10}
\Delta z = \frac{z_{\rm \max} - z_{\rm \min}}{M}.
\end{equation}

\noindent
Group $i$ is represented by $z_i$, which is

\begin{equation}
\label{eq11}
z_i = z_{\rm \min} + \left(i - \frac12 \right)\Delta z, ~i = 1, 2, .., M - 1, M.
\end{equation}

\subsection{Member of each group}

As it has been declared previously, in group $i$ there is only one value of $z$ which is $z_i$. It is only for the sake of simplicity. Each group has number of particles that must obey the Gaussian distribution function. Number of particle in each group is

\begin{equation}
\label{eq12}
N_i = \left(\frac{N}{N'}\right) {\rm int}[N(z_i)].
\end{equation}

\noindent
Since there is a round down process (throug the {\tt int()} function) for each group in order to find $N_i$ from $N(z_i)$ then it can be concluded that

\begin{equation}
\label{eq13}
\sum_{i = 1}^M N_i \le N,
\end{equation}

\noindent
a difference that deviates the discreet groups of particles from the Gaussian distribution function. The factor in front of right side of Equation (\ref{eq12}) is due to discreet number of particle groups.

\subsection{Algorithm to group the particles}

An algorithm of implementation of Equation (\ref{eq08}) - (\ref{eq12}) can be summerized as follow

\begin{enumerate} \tt
\item{\label{st11}} start
\item{\label{st12}} determine mu and sigma for distribution function N(z)
\item{\label{st13}} determine epsilon
\item{\label{st14}} set z = mu
\item{\label{st15}} using root finding algoritm find root of N(z) - (1 - epsilon) = 0 in range z < mu, it is named as zmin
\item{\label{st16}} set z = mu
\item{\label{st17}} using root finding algoritm find root of N(z) - (1 - epsilon) = 0 in range z > mu, it is named as zmax
\item{\label{st18}} determine number of group M
\item{\label{st19}} calculate group width dz using Equation (\ref{eq10})
\item{\label{st1A}} determine zi using Equation (\ref{eq11}) for all M groups
\item{\label{st1B}} deterimine number of group i using Equation (\ref{eq12})
\item{\label{st1C}} calculate N' and normalize Ni with it
\item{\label{st1D}} stop
\end{enumerate}

\section{The sequences}

In group $i$ there are $N_i$ particles which has a property $z_i$. The property can be velocity, mass, diameter, charge, or other physical properties. And there are $M$ groups of particles. It means, when all the particles are lined in order to make sequences there will be $S$ ways to rearrange the particles order. If the particles are distinguishable

\begin{equation}
\label{eq14}
S_{\rm distinguishable} = N!
\end{equation}

\noindent
and when there are indistinguishable

\begin{equation}
\label{eq15}
S_{\rm indistinguishable} = \frac{N!}{\prod_{i = 1}^M N_i!}.
\end{equation}

\noindent
The later means that particles at the same group are identical, which means the particles are identify only by their property $z_i$.

\subsection{The zeroth sequence}

The easiest way to buid the sequence is by lining the particle from each group in incremental order, such as

\begin{equation}
\label{eq16}
z_1, z_1, z_2, z_2, z_2, z_2, z_3, z_3, .., z_M, z_M.
\end{equation}

\noindent
This sequence is named as the zeroth sequence.

\subsection{Other sequences}

The sequences beside zeroth sequence can be generated by permutating zeroth sequence. Number of sequences can be produced is according to Equation (\ref{eq14}) and (\ref{eq15}). In this report we propose a mechanism to generate a sequence from zeroth sequence by using {\tt random()} and {\tt swap()} function which is already built-in in C++. The algorithm is as follow

\begin{enumerate} \tt
\item{\label{st21}} start
\item{\label{st22}} determine seed for random generator
\item{\label{st23}} set the generator with the seed
\item{\label{st24}} get the zeroth sequence that contains N particles
\item{\label{st25}} particle number i = 1
\item{\label{st26}} generate an integer number between 1 and N, say j
\item{\label{st27}} swap value of particle i and j
\item{\label{st28}} increase value of i by 1
\item{\label{st29}} if i still less than or equal to M go to Step \ref{st26}
\item{\label{st2A}} stop
\end{enumerate}

\noindent
Since random number generated by C++ random generator depends on the seed, than the sequence is reproducible. It means that the seed is as an idenfier to the sequence.

\section{Error}

The sum of generated value of $N_i$ for each group $i$ in a sequence will be less than total number of particle as given by Equation (\ref{eq13}), which means an error. This error can be calculated using a common correlation coefficient $R^2$ formulation

\begin{equation}
\label{eq17}
R^2 = 1 - \frac{SS_{\rm err}}{SS_{\rm tot}},
\end{equation}

\noindent
where

\begin{eqnarray}
\label{eq18}
SS_{\rm err} = \sum_i [N_i - N(z_i)]^2, \\
\label{eq19}
SS_{\rm tot} = \sum_i (N_i - \overline{N}_i)^2, \\
\label{eq20}
\overline{N}_i = \frac{1}{N'} \sum_i N_i,
\end{eqnarray}

\noindent
with $N'$ is total number of generated particles

\begin{equation}
\label{eq21}
N' = \sum_{i = 1}^M N_i.
\end{equation}

\noindent
Equation (\ref{eq17}) -  (\ref{eq21}) will be used the next section to calculate the error in produced sequences.

\section{Results and discusion}

\begin{figure}[h]
\centering
\includegraphics[width=9cm]{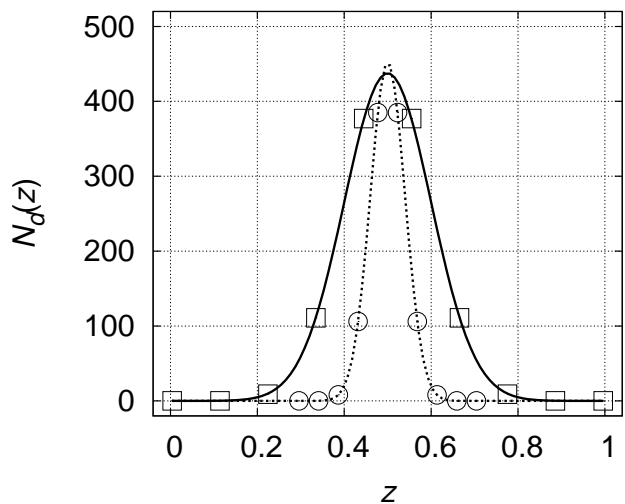}
\caption{\label{fg02} Example of discreet value of Gaussian distribution function generated by {\tt gaussg} with $\mu = 0.5$ for $\sigma = 0.1$, $N' = 994$, $N_d = 109.665$ (solid line and square mark) and $\sigma = 0.04$, $N' = 998$, $N_d = 45.3348$ (dashed line and circle mark).}
\end{figure}

An illustration for two discreet Gaussian distribution function is given in Figure \ref{fg02}, which is produced by our program {\tt gaussg}. It has been found that the value $N'$ shown in Equation (\ref{eq12}) can not be used in the continue function to fit the discreet values. Then, the new fitting function will be

\begin{equation}
\label{eq22}
N_d(z) = \frac{N_d}{\sigma\sqrt{2\pi}} \exp\left[-\frac{(z - \mu)^2}{2\sigma^2}\right],
\end{equation}

\noindent
where

\begin{equation}
\label{eq23}
N_d = \frac{N N'}{\sum_{i = 1}^M N(z_i)}.
\end{equation}

\noindent
The correlation coefficient in Equation (\ref{eq17}) is caculated using $N_d(z_i)$ instead of $N(z_i)$.

\begin{figure}[h]
\centering
\includegraphics[width=9cm]{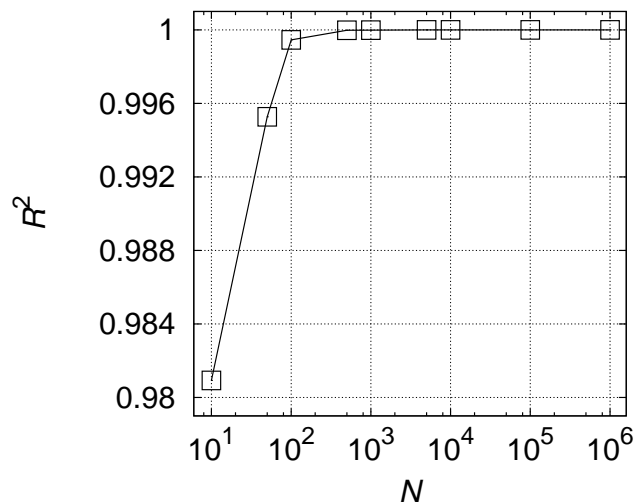}
\caption{\label{fg03} Dependence of correlation coefficient $R^2$ on number of particles $N$ for $\mu = 0.5$, $\sigma = 0.1$, and $M = 10$.}
\end{figure}

\begin{figure}[h]
\centering
\includegraphics[width=9cm]{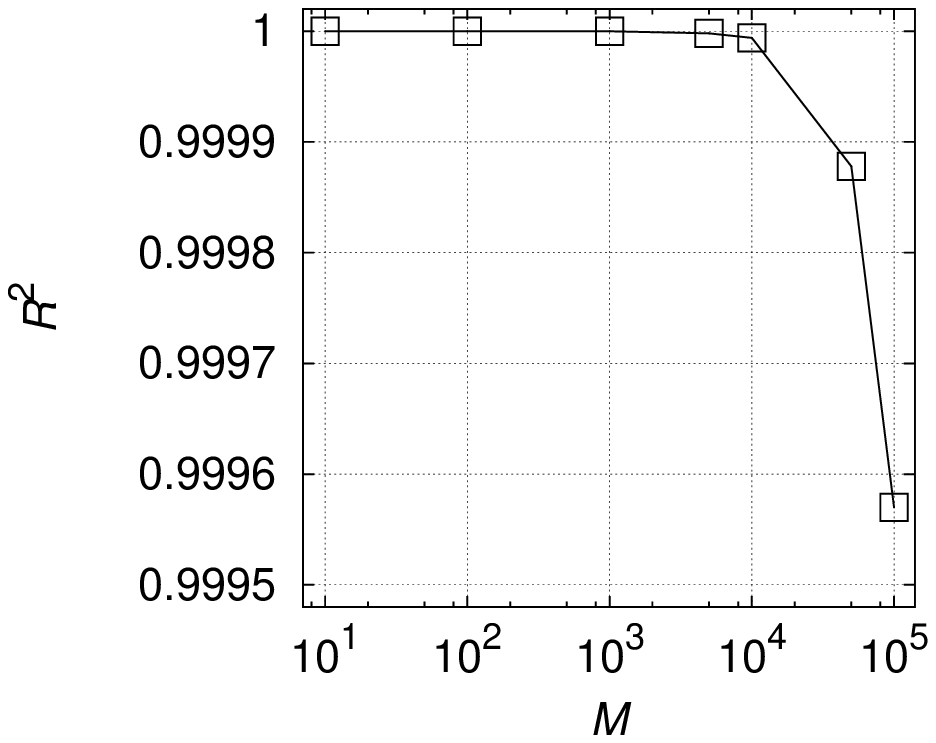}
\caption{\label{fg04} Dependence of correlation coefficient $R^2$ on number of groups $M$ for $\mu = 0.5$, $\sigma = 0.1$, and $N = 1000000$.}
\end{figure}

\medskip
Variation of number of particles $N$ and number of groups $M$ are also observed as illustrated in Figure \ref{fg03} and Figure \ref{fg04}, respectively. It can be seen that larger $N$ gives better $R^2$ and larger $M$ gives bad $R^2$. Number of groups should be more than or equal to $N / M$ that the program {\tt gaussg} can handled.

\begin{figure}[h]
\centering
\includegraphics[width=12cm]{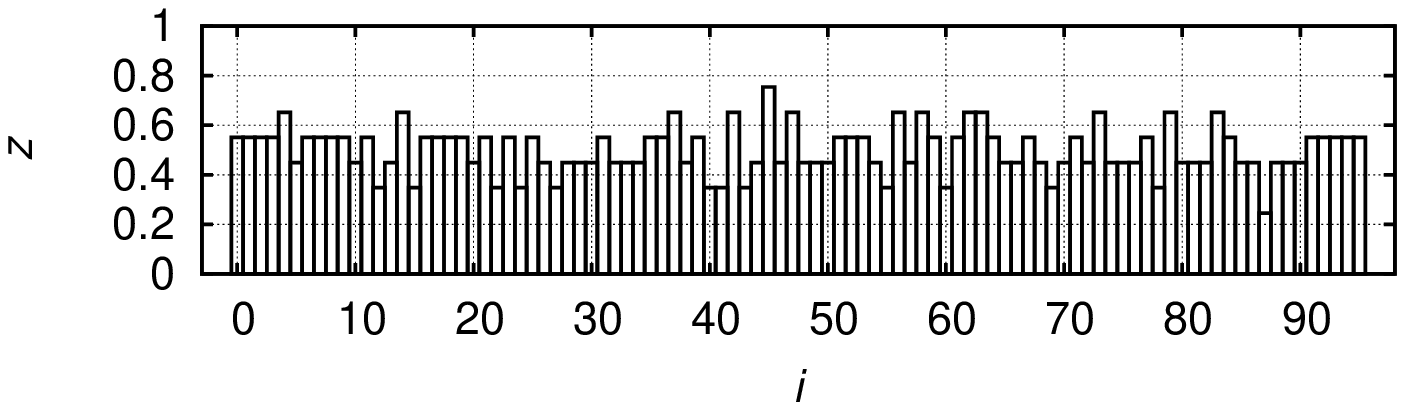} \\
(a) \\
\includegraphics[width=12cm]{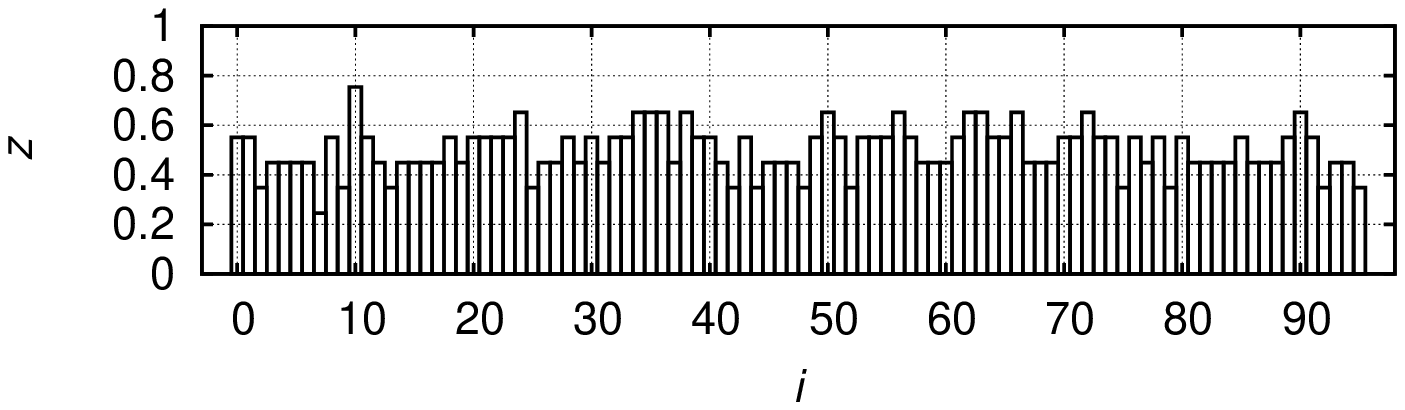} \\
(b) \\
\includegraphics[width=12cm]{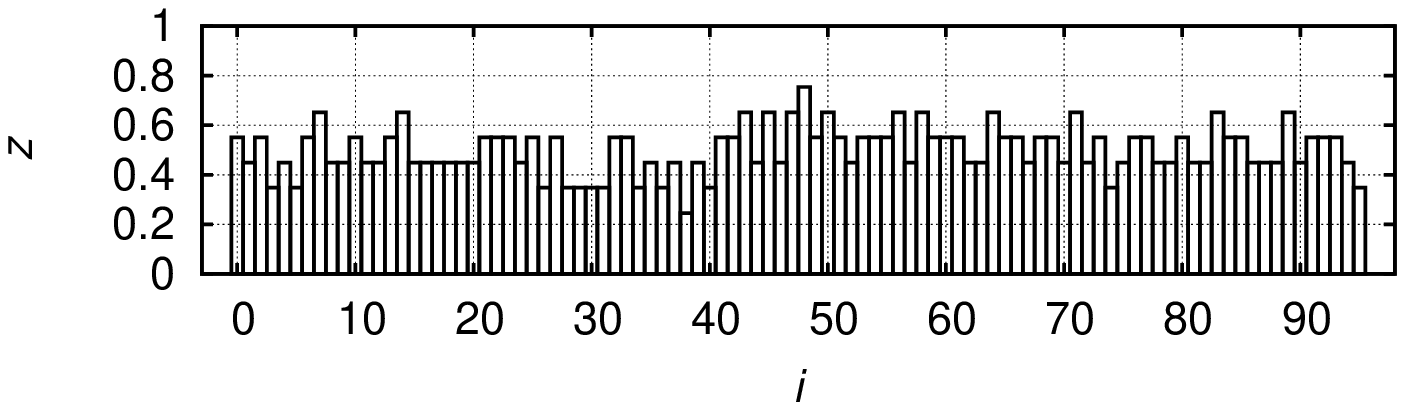} \\
(c) \\
\includegraphics[width=12cm]{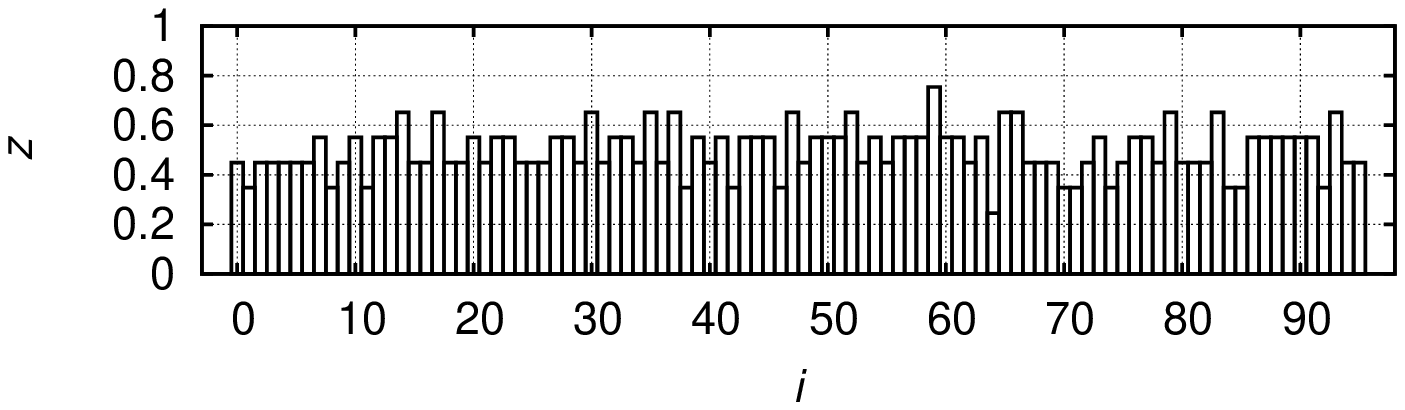} \\
(d)
\caption{\label{fg05} Sequences with seed: (a) 1, (b) 2, (c) 3, and (d) 4.}
\end{figure}

\medskip
The next results are the sequences that produced from $N_d(z_i)$ as shown in Figure \ref{fg05}. Only first four seeds are used to generate four sequences. These sequences has the same distribution function, which has $\mu = 0.5$,, $\sigma = 0.1$, $N = 100$, and $M = 10$. These results are produced by program {\tt gausss}.

\section{Conclusion}

Two programs, {\tt gaussg} for creating discreet groups and {\tt gausss} for creating sequences, have been devoleped and tested. The discreen Gaussian distribution function can be produced. The sequences which has the same distribution function, can also be generated.  Further investigation is needed how to register all available sequences for a distribution function. As $N$ increases the value $R^2$ approximates 1, but as $M$ increases the value $R^2$ decrease less than 1. $R^2 = 1$ can be achieved with larger $N$ and smaller $M$. The discreet Gaussian distribution function has different constant with its previously continuos distribution function which is used to generated the discreet and limited groups.

\bigskip\bigskip\noindent
{\bf \Large Acknowledgements}\\ \\
Authors would like to thank to Institut Teknologi Bandung Alumni Association Research Grant in year 2011 for partially supporting to this work.

\bibliographystyle{unsrt}
\bibliography{manuscript}

\bigskip\bigskip\noindent
{\bf \Large Appendix A: gaussg}\\ \\

{\small
\begin{verbatim}
/*
    gaussg.cpp
    Generate discreet groups of Gaussian distribution function
    Authors are Sparisoma Viridi and Veinardi Suendo
    Version date is 2011.07.17
*/

#include <iostream>
#include <fstream>
#include <stdlib.h>
#include <math.h>

const double PI = 3.14159265;

using namespace std;

double Nz(double mu, double sigma, double N, double z);

int main(int argc, char **argv) {
    if(argc < 6) {
        cout << "Version date is 2011.07.17" << endl;
        cout << "gaussg is written by Sparisoma Viridi ";
        cout << "and Veinardi Suendo" << endl;
        cout << "Generate discreet groups of particles ";
        cout << "that obey Gaussian distribution ";
        cout << "fuction" << endl;
        cout << endl;
        cout << "Usage: gaussg mu sigma N M output-file" << endl;
        cout << endl;
        cout << "All arguments are mandatory:" << endl;
        cout << "mu           average of Gaussian ";
        cout << "distribution function" << endl;
        cout << "sigma        width of Gaussian ";
        cout << "distribution function" << endl;
        cout << "N            number of particles" << endl;
        cout << "M            number of groups" << endl;
        cout << "output-file  output file" << endl;
    } else {
        double mu = atof(argv[1]);
        double sigma = atof(argv[2]);
        int N = atoi(argv[3]);
        int M = atoi(argv[4]);
        const char *ofn = argv[5];
        
        cout << "mu = " << mu << endl;
        cout << "sigma = " << sigma << endl;
        cout << "N = " << N << endl;
        cout << "M = " << M << endl;
        cout << "output-file = " << ofn << endl;
        
        double eps = 1E-3;
        double dz = mu * 1E-5;
        double NNz = N;
        
        double zmin = mu;
        while(NNz > eps) {
            NNz = Nz(mu, sigma, N, zmin);
            zmin -= dz;
        }
        cout << "zmin = " << zmin << endl;
        
        NNz = N;
        double zmax = mu;
        while(NNz > eps) {
            NNz = Nz(mu, sigma, N, zmax);
            zmax += dz;
        }
        cout << "zmax = " << zmax << endl;
        
        dz = (zmax - zmin) / M;
        cout << "dz = " << dz << endl;
        
        double zi[M];
        int Ni[M];
        double NN = 0;
        for(int i = 0; i < M; i++) {
            zi[i] = zmin + (i + 0.5) * dz;
            double z = zi[i];
            Ni[i] = (int) Nz(mu, sigma, N, z);
            NN += Ni[i];
        }
        cout << "N' = " << NN << endl;
        
        double NN2 = 0;
        for(int i = 0; i < M; i++) {
            Ni[i] = (int)(Ni[i] * (N/NN));
            // cout << i << "\t";
            // cout << Ni[i] << endl;
            NN2 += Ni[i];
        }
        cout << "N\" = " << NN2 << endl;
        
        double Nzi[M];
        
        double NN3 = 0;
        ofstream fout;
        fout.open(ofn);
        fout << "#i\tzi\tNi\tN(zi)" << endl;
        for(int i = 0; i < M; i++) {
            fout << i + 1 << "\t";
            fout << zi[i] << "\t";
            fout << Ni[i] << "\t";
            double z = zi[i];
            NN3 = 1.0 * N * NN2 / NN;
            Nzi[i] = Nz(mu, sigma, NN3, z);
            fout << Nzi[i] << endl;
        }
        fout.close();
        cout << "Nt = " << NN3 << endl;
        
        double SNi = 0;
        for(int i = 0; i < M; i++) {
            SNi += (Ni[i] * zi[i]);
        }
        double mui = SNi / NN2;
        
        double SStot = 0;
        double SSerr = 0;
        for(int i = 0; i < M; i++) {
            double dSStot = (Ni[i] - mui) * (Ni[i] - mui);
            SStot += dSStot;
            double dSSerr = (Ni[i] - Nzi[i]) * (Ni[i] - Nzi[i]);
            SSerr += dSSerr;
        }
        double R2 = 1 - SSerr/SStot;
        cout << "R^2 = " << R2 << endl;
    }
    return 0;
}

double Nz(double mu, double sigma, double N, double z) {
    double c1 = N  / (sigma * sqrt(2 * PI));
    double c2 = exp(-(z - mu)*(z - mu) / (2 * sigma * sigma));
    double c3 = c1 * c2;
    return c3;
}
\end{verbatim}
}

\bigskip\bigskip\noindent
{\bf \Large Appendix B: gausss}\\ \\

{\small
\begin{verbatim}
/*
    gausss.cpp
    Generate sequences from discreet groups of Gaussian
    distribution function
    Authors are Sparisoma Viridi and Veinardi Suendo
    Version date is 2011.07.17
*/

#include <iostream>
#include <fstream>
#include <stdlib.h>
#include <math.h>

const double PI = 3.14159265;

using namespace std;

int main(int argc, char **argv) {
    if(argc < 4) {
        cout << "Version date is 2011.07.17" << endl;
        cout << "gausss is written by Sparisoma Viridi ";
        cout << "and Veinardi Suendo" << endl;
        cout << "Generate sequences from discreet groups ";
        cout << "of particles that\nobey Gaussian ";
        cout << "distribution fuction" << endl;
        cout << endl;
        cout << "Usage: seed input-file output-file" << endl;
        cout << endl;
        cout << "All arguments are mandatory:" << endl;
        cout << "seed         seed for random generator ";
        cout << "(1, 2, ..)";
        cout << endl;
        cout << "input-file   input file" << endl;
        cout << "output-file  output file" << endl;
    } else {
        long int seed = atoi(argv[1]);
        const char *ifn = argv[2];
        const char *ofn = argv[3];
        double mu = atof(argv[1]);
        
        cout << "seed = " << seed << endl;
        cout << "input-file = " << ifn << endl;
        cout << "output-file = " << ofn << endl;
        
        ifstream fin;
        fin.open(ifn);
        string buf;
        double d;
        long int i = 0;
        while(!fin.eof()) {
            fin >> buf;
            i++;
        }
        fin.close();
        
        int M = (int)((i - 1 - 4) / 4);
        double zi[M], Nzi[M];
        
        fin.open(ifn);
        int j = 0;
        fin >> buf; fin >> buf; fin >> buf; fin >> buf;
        while(!fin.eof() || j < i) {
            int k; fin >> k;
            fin >> zi[k-1];
            fin >> Nzi[k-1];
            fin >> buf;
            j++;
        }
        
        int N = 0;
        for(int l = 0; l < M; l++) {
            zi[l] = 0.001 * round(zi[l] * 1000);
            N += Nzi[l];
        }
        
        double seq0[N];
        int k = 0;
        for(int m = 0; m < M; m++) {
            for(int l = Nzi[m]; l > 0; l--) {
                seq0[k] = zi[m];
                k++;
            }
        }
        
        double seq1[N];
        for(int n = 0; n < N; n++) {
            seq1[n] = seq0[n];
        }
        
        srandom(seed);
        for(int n = 0; n < N; n++) {
            long int a1 = random();
            double a2 = 1.0 * a1 / RAND_MAX;
            int a3 = (int)(N * a2);
            swap(seq1[n], seq1[a3]);
        }
        
        ofstream fout;
        fout.open(ofn);
        for(int n = 0; n < N; n++) {
            fout << seq1[n] << endl;
        }
        fout.close();
    }
    return 0;
}
\end{verbatim}
}

\end{document}